\documentclass[letterpaper]{article} 
\usepackage{aaai2026}  
\usepackage{times}  
\usepackage{helvet}  
\usepackage{courier}  
\usepackage[hyphens]{url}  
\usepackage{graphicx} 
\usepackage{natbib}  
\usepackage{caption} 
\usepackage{colortbl}
\definecolor{lightred}{rgb}{1, 0.8, 0.8}
\definecolor{lightgreen}{rgb}{0.8, 1, 0.8}
\definecolor{lightblue}{rgb}{0.8, 0.8, 1}
\definecolor{lightyellow}{rgb}{1, 1, 0.8}
\usepackage{makecell}
\usepackage{multirow}
\usepackage{algorithm}
\usepackage{algorithmic}
\usepackage{adjustbox}
\usepackage{newfloat}
\usepackage{listings}
\DeclareCaptionStyle{ruled}{labelfont=normalfont,labelsep=colon,strut=off} 
\floatstyle{ruled}
\newfloat{listing}{tb}{lst}{}
\floatname{listing}{Listing}
\title{Bridging Brains and Models: MoE-Based Functional Lesions for Simulating and Rehabilitating Aphasia}
\author{
    Yifan Wang\textsuperscript{\rm 1}\equalcontrib,
    Jingyuan Sun\textsuperscript{\rm 1}\equalcontrib,
    Jichen Zheng\textsuperscript{\rm 2,\rm 3},
    Yunhao Zhang\textsuperscript{\rm 2,\rm 3},\\
    Chunyu Ye\textsuperscript{\rm 2,\rm 3},
    Jixing Li\textsuperscript{\rm 4},
    Chengqing Zong\textsuperscript{\rm 2,\rm 3},
    Shaonan Wang\textsuperscript{\rm 2,\rm 3}\corres
}
\affiliations{
    \textsuperscript{\rm 1}Department of Computer Science, The University of Manchester, UK,\\
    \textsuperscript{\rm 2}State Key Laboratory of Multimodal Artificial Intelligence Systems, Institute of Automation, CAS, Beijing, China,\\
    \textsuperscript{\rm 3}School of Artificial Intelligence, University of Chinese Academy of Sciences, Beijing, China,\\
    \textsuperscript{\rm 4}Department of Linguistics and Translation, City University of Hong Kong, Hong Kong\\
    yifan.wang-38@postgrad.manchester.ac.uk\\
    jingyuan.sun@manchester.ac.uk\\
    \{zhengjichen2023, zhangyunhao2021, yechunyu2023\}@ia.ac.cn\\
    jixingli@cityu.edu.hk\\
    \{shaonan.wang, cqzong\}@nlpr.ia.ac.cn
}

\begin{document}

\maketitle

\renewcommand{\thefootnote}{\fnsymbol{footnote}}

\footnotetext[1]{Corresponding author.} 
\footnotetext[2]{These authors contributed equally.} 

\renewcommand{\thefootnote}{\arabic{footnote}}

\begin{abstract}
The striking alignment between large language models (LLMs) and human brain activity positions them as powerful models of healthy cognition. This parallel raises a fundamental question: if LLMs can model the intact brain, can we lesion them to simulate the linguistic deficits of the injured brain? In this work, we introduce a methodology to model aphasia —a complex language disorder caused by neural injury— by selectively disabling components in a modular Mixture-of-Experts (MoE) language model. We simulate distinct aphasia subtypes, validate their linguistic outputs against real patient speech, and then investigate functional recovery by retraining the model's remaining healthy experts. Our results demonstrate that lesioning functionally-specialized experts for syntax or semantics induces distinct impairments that closely resemble Broca’s and Wernicke’s aphasia, respectively. Crucially, we show that freezing the damaged experts and retraining the intact ones on conversational data restores significant linguistic function, demonstrating a computational analogue for rehabilitation. These findings establish modular LLMs as a powerful and clinically-relevant potential framework for modeling the mechanisms of language disorders and for computationally exploring novel pathways for therapy.
\end{abstract}

\begin{figure}[ht]
    \centering
    \includegraphics[width=0.34\textwidth]{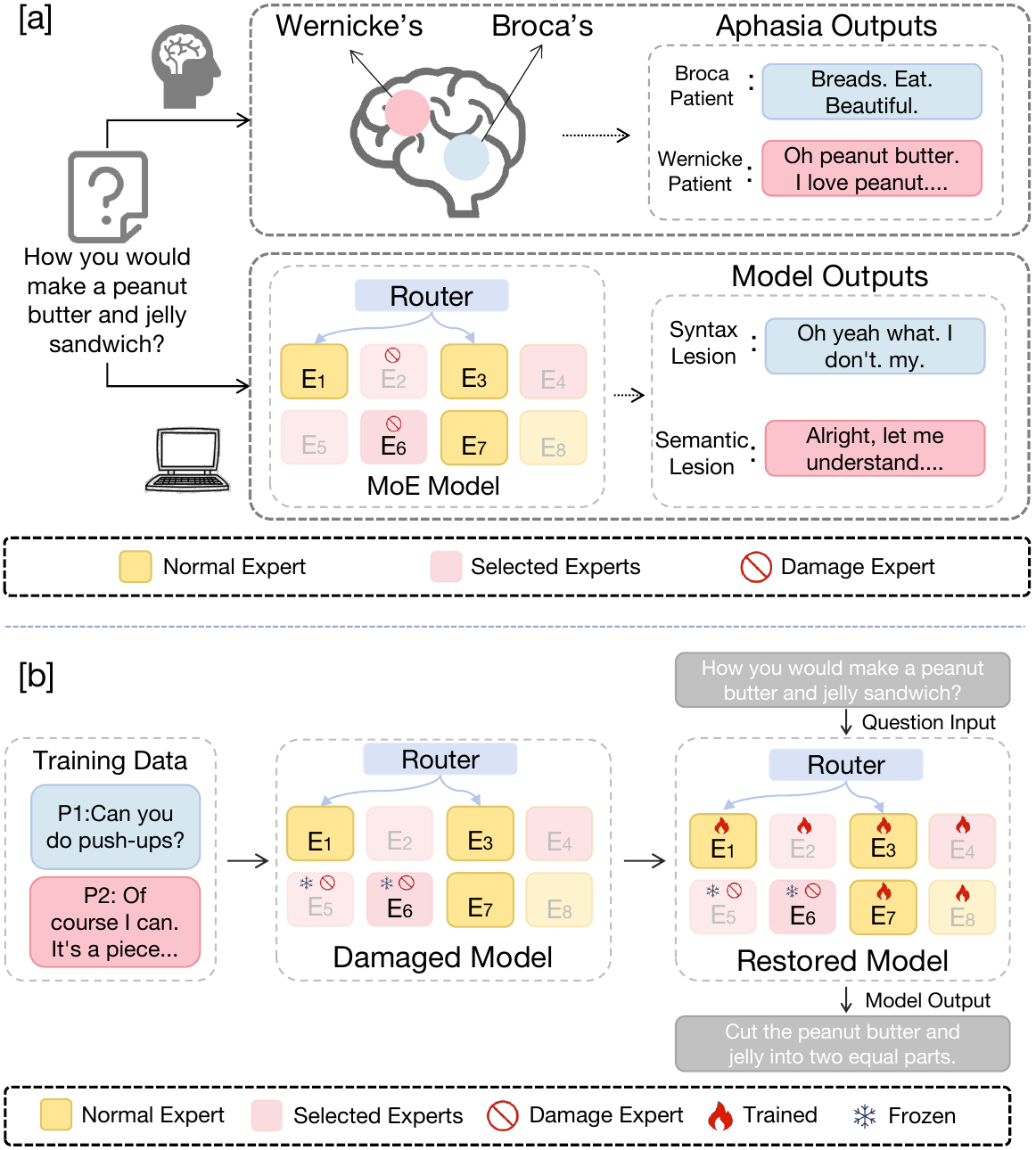}
    \caption{\textbf{Overview of the analysis pipeline.} [a] Aphasia Simulation via Expert Lesioning: We selectively damage experts within a MoE model to induce linguistic impairments. The model's outputs are then compared to the characteristic speech patterns of patients with Broca's and Wernicke's aphasia. [b] Rehabilitation via Retraining: We model a therapeutic process by freezing the damaged experts and fine-tuning the remaining intact experts, testing for the recovery of language function.}
    \label{fig:1}
\end{figure}

\section{Introduction}

Beyond their impressive performance, large language models (LLMs) are increasingly recognized for their ability to simulate a wide range of human behaviors \citep{park2023generative, mou2024individual}. Concurrently, research has established a striking alignment between their internal representations and the neural activity in the human brain during language processing \citep{schrimpf2021neural, caucheteux2022brains,toneva2019interpreting, goldstein2024alignment}. This alignment with healthy, typical cognition raises a compelling question: if LLMs can partially model the intact brain, can we lesion them to simulate the impaired brain? Specifically, can we manipulate an LLM's components to model the distinct linguistic behaviors of patients with aphasia, a complex language disorder resulting from neural injury? Success in this endeavor would yield a highly promising tool: a computational proxy for patients, enabling researchers to test therapeutic hypotheses in a controlled, scalable environment. Furthermore, it would provide a novel paradigm for investigating the functional architecture of both the human brain and the artificial models that seek to emulate it in language processing.

The scientific foundation for such an approach lies in the modular organization of language in the brain, a principle established over decades of neuroscience research \citep{dronkers2023neuroscience, meunier2010modular, bertolero2015modular}. This framework posits that distinct brain regions are specialized for different linguistic functions, and damage to these regions results in predictable deficits. For instance, Broca’s and Wernicke’s aphasias, two of the most prevalent subtypes, are classically associated with damage to regions governing speech production and comprehension, respectively. Early attempts to model these phenomena used small-scale connectionist models, demonstrating that lesioning specific components could replicate certain patient behaviors \citep{dell1997lexical, farah1991cognitive, hinton1991lesioning}. While foundational, these pioneering models lacked the scale and linguistic sophistication required to capture the full complexity of human language, limiting their utility as realistic proxies for aphasic syndromes.

To bridge this gap, we leverage a modern Mixture-of-Experts (MoE) language model, whose architecture theoretically offers a compelling architectural parallel. Research confirms that individual 'experts' in MoE models become functionally specialized for distinct tasks and linguistic phenomena \citep{chen2022task, zhang2023emergent}. This emergent modularity makes them a uniquely suitable framework for simulating the effects of localized brain damage. In this work, we use the OLMoE model \citep{muennighoff2024olmoe} to investigate aphasia through a multi-stage process of simulation, validation, and rehabilitation, as shown in Figure 1. First, we systematically lesion individual experts within the model to simulate impairments of varying type and severity, identifying the functional contributions of each expert by evaluating performance on semantic and syntactic tasks. Second, to validate the clinical authenticity of our simulations, we compare the model's linguistic outputs directly against real speech samples from aphasia patients. Finally, we model a therapeutic process by freezing the lesioned experts and retraining the remaining components, testing whether the intact modules can compensate for lost function and restore linguistic abilities.

Our work makes several key contributions. We demonstrate that targeted lesioning of an MoE model can successfully replicate the distinct linguistic profiles of Broca's and Wernicke's aphasia. Our findings reveal a novel mapping: Broca’s-like deficits correlate with widespread, distributed expert damage, whereas Wernicke’s-like deficits arise from more localized expert failure. Critically, we show that the remaining healthy experts can be retrained to partially recover lost function, offering a proof-of-concept for using computational models to explore and design personalized rehabilitation strategies. Together, these results establish MoE models' potential as a viable and powerful new tool for computational neuropsychology.

\section{Related Work}

Our research integrates insights from three distinct but converging domains: the neuroscience of aphasia, the history of computational disorder simulation, and the architecture of modern language models.

\subsection{Aphasia and the Neural Basis of Language}

Aphasia, a language disorder resulting from brain damage, provides a crucial window into the neural organization of language. The classic distinction between Broca's aphasia, characterized by effortful, agrammatical speech but preserved comprehension, and Wernicke's aphasia, marked by fluent but semantically incoherent speech, offers strong evidence for a functional dissociation between syntactic and semantic processing \citep{lichtheim1885aphasia, damasio1992aphasia}. Decades of neuroimaging and lesion studies have reinforced this modular view. The left inferior frontal gyrus (IFG) is consistently implicated in syntactic processing \citep{friederici2011brain}, while regions in the temporal lobe, such as the middle temporal gyrus (MTG) and anterior temporal lobe (ATL), are critical for semantic integration \citep{dronkers2004lesion, friederici2011brain}. This established link between localized brain damage and specific linguistic deficits provides the theoretical bedrock for our approach: modeling aphasia by selectively impairing components in a neural language model.

\subsection{Simulating Language Disorders with Computational Models} 

The effort to simulate neurological disorders has a long history in cognitive science. Early connectionist models successfully replicated naming errors and grammatical deficits by lesioning network components, establishing the viability of the approach \citep{joanisse1999impairments, thomas2002modelling}. With the advent of large language models (LLMs), researchers have gained more powerful tools. Recent studies have used LLMs to simulate aphasic speech through mechanisms like controlled sentence completion \citep{manir2024llm} or by damaging layers in multimodal models to induce specific deficits \citep{wang2025emergent}. While these studies demonstrate the potential of modern architectures, they often involve lesioning monolithic models where the link between a damaged component and a specific linguistic function is not always clear. Our work builds on this tradition but seeks a model with a more direct and biologically plausible analogy to the brain's modularity.

\subsection{Mixture-of-Experts Models and Functional Modularity} Mixture-of-Experts (MoE) models offer a compelling architectural solution. Originally developed to scale models efficiently by activating only a subset of parameters ("experts") for any given input \citep{shazeer2017outrageously}, the MoE paradigm has become central to state-of-the-art LLMs like Switch Transformer and Mixtral \citep{fedus2022switch, jiang2024mixtral}. Beyond efficiency, a key property of MoE models is the emergence of functional specialization. Research shows that individual experts develop preferences for specific tasks or domains \citep{chen2022task}, and recent work provides direct evidence that different experts become responsible for distinct linguistic phenomena \citep{zhang2023emergent}. This inherent modularity, where specialized sub-networks contribute to the final output, makes MoE models a uniquely suitable framework for simulating the effects of localized brain damage. Our work is situated at the intersection of these fields. While prior work has lesioned language models, we are the first to leverage the emergent functional modularity of MoE models to conduct expert-level lesion studies of aphasia. Crucially, by not only simulating deficits and validating them against patient data from AphasiaBank but also modeling functional recovery via retraining, we move beyond mere simulation to explore computational analogues of rehabilitation.

\begin{figure*}[t]
    \centering
    \includegraphics[width=0.96\textwidth]{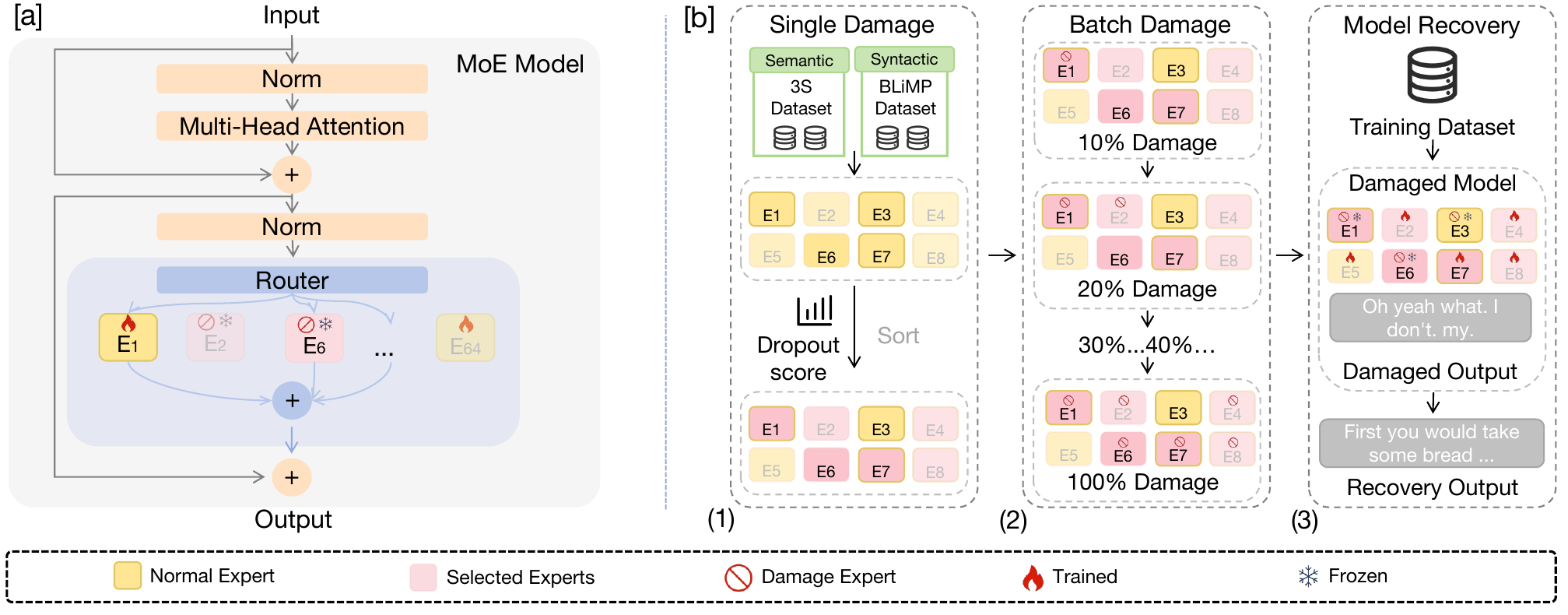}
    \caption{\textbf{Details of the analysis pipeline.} [a] Model structure of the OLMoE with a part of experts lesioned \citep{muennighoff2024olmoe}. [b] The three-stage experimental framework:  (1)  We lesion experts individually and measure performance drops on the 3S (semantic) and BLiMP (syntactic) datasets to identify their functional roles. (2) We then batch-lesion 10\% to 100\% of these critical experts to simulate varying impairment severity. (3) Finally, we model functional recovery by retraining the damaged model on a therapy corpus, while keeping the lesioned experts frozen.}
    \label{fig:fig2}
\end{figure*}

\section{Method}
Our framework is built on a transformer-based Mixture-of-Experts (MoE) model and consists of three steps: (1) analyzing the functionality of individual experts, (2) simulating aphasia-like language output through expert ablation, and (3) retraining the model to simulate recovery (Shown in Figure \ref{fig:fig2}[b]).In the following sections, we will first briefly introduce the MoE model architecture, then detail how we investigate expert functions, simulate aphasic language patterns, and retrain the model for recovery.

\subsection{MoE Model Architecture}

 We adopt the OLMoE model proposed by Muennighoff et al. (2024) \citep{muennighoff2024olmoe}, specifically the OLMoE-1B-7B-0924-Instruct version. As shown in Figure \ref{fig:fig2}[a], OLMoE follows the standard Transformer architecture, with the Feedforward Network (FFN) layers replaced by MoE layers. A lightweight linear gating network (Router) selects 8 out of 64 experts per token. The model contains 16 layers, with a total of 1B activated parameters, and is fine-tuned on OLMo-Instruct data.

\subsection{MoE Model Expert Damage}
In order to simulate the permanent damage of a specific brain region with the model, we chose to intervene in the expert module by destroying its internal weights. This approach ensures that the expert module can no longer use its original learned knowledge, but maintains the integrity of the network structure. We use Xavier uniform initialization to reset the weights of the selected expert module. Xavier initialization is a standard and widely verified weight initialization technique, whose goal is to keep the variance of activation values and gradients in each layer of the network stable \citep{glorot2010understanding}. In the expert damage step, the selected expert module in the model is traversed and all its weight parameters are overwritten with new values sampled from the Xavier uniform distribution. The formula is shown in Formula 1:
\begin{equation}
a = \sqrt{\frac{6}{n_{in} + n_{out}}}
\end{equation}

$n_{in}$ represents the number of input neurons of the expert module, and $n_{out}$ represents the number of output neurons. We overwrite the original weights with a random value with a mean of 0 and a moderate variance, thereby completely destroying the original function of the expert module and rendering it unable to perform its specific task.

We adopted Xavier initialization to impair expert modules, rather than setting weights to zero or assigning arbitrary random values, based on two key considerations: (1) Zeroing the weights completely silences the expert’s output, eliminating any signal during forward propagation—an unrealistic representation of brain injury, where regions often remain active despite functional loss. (2) Using uncalibrated random values risks disrupting signal variance, potentially introducing numerical instability during training or inference. Xavier initialization offers a principled randomization scheme that erases learned functionality while preserving stable signal flow. This ensures that the damaged expert remains structurally active but functionally disrupted—closely mimicking the neurological condition where damaged brain regions exhibit preserved metabolic activity despite impaired functionality \citep{kiran2019neuroplasticity, gleichgerrcht2015preservation, wilson2020neuroplasticity}.

\subsection{Exploring functionality of individual experts}
To simulate distinct aphasia subtypes, we selectively damage expert modules within a Mixture-of-Experts (MoE) language model. Our methodology targets experts most crucial to semantic and syntactic processing, enabling controlled simulation of aphasic deficits. Wernicke’s aphasia, marked by semantic deficits, is modeled by lesioning experts critical for meaning; Broca’s aphasia, involving syntactic impairments, is replicated by targeting experts tied to grammar.

To identify these task-relevant experts, we employ four datasets aligned with the two language components:
\begin{itemize}
    \item \textbf{Semantics:} The 3S suite (SICK, STS-B, STS12) for sentence-level semantic similarity.
    \item \textbf{Syntax:} BLiMP-Syntax for fine-grained syntactic judgments.
\end{itemize}

Initially, we evaluate the intact model across all datasets to establish baseline performance metrics. Subsequently, we perform systematic per-layer, per-expert ablations, damaging one expert at a time and evaluating the resulting model on all tasks. After each evaluation, we restore the original model configuration, ensuring independent expert assessments. This generates a detailed mapping of individual experts to semantic or syntactic performance.

We then identify experts whose ablation consistently reduces task performance. These experts are ranked according to the severity of their impact, creating an importance-based hierarchy. Recognizing functional redundancy and interactions among experts, we employ a hierarchical cumulative ablation strategy to simulate varying degrees of language impairment. Specifically, we incrementally ablate experts grouped by importance tiers (top 10\%, 20\%, …, up to 100\% of affected experts) and re-assess performance at each cumulative stage.

This structured approach enables analysis of nonlinear expert interactions and quantifies how language performance deteriorates with progressive expert loss. Ultimately, we identify expert subsets capable of accurately simulating mild to severe aphasic symptoms, providing a robust, computational framework to model and investigate different intensities of Wernicke’s and Broca’s aphasia.

\subsection{Simulating aphasia-like language generation}
Following expert ranking and batch ablation, we constructed two targeted model variants: one simulating Wernicke’s aphasia by lesioning experts involved in semantic processing, and another simulating Broca’s aphasia by damaging those involved in syntactic generation. These variants aim to emulate the functional language deficits observed in real aphasia patients.

To validate the plausibility of these simulations, we conducted an evaluation using real-world conversational data from the AphasiaBank corpus, which includes clinical interviews with patients diagnosed with various types of aphasia and healthy controls. We selected three groups: 
\begin{itemize}
    \item \textbf{Wernicke's aphasia patients:} Typical features are confusion of semantics, vocabulary errors, fluent speech but lack of information;
    \item \textbf{Broca's aphasia patients:} Simplified grammatical structure, difficulty in pronunciation, unfluent expression but clear semantics;
    \item \textbf{Normal control group:} Healthy subjects without language disorders, used for comparative analysis.
\end{itemize}

For each sample, interviewer prompts (INV) were extracted and used as model inputs across three variants: the undamaged model (baseline), the Wernicke simulation model, and the Broca simulation model. The model-generated responses (PAR) were then compared to real patient responses using several metrics:
\begin{itemize}
    \item Sentence-level semantic similarity calculation (such as Cosine Similarity)
    \item Language complexity analysis (such as vocabulary richness)
    \item Structural feature matching (such as subject-verb-object integrity, error rate, missing function words)
\end{itemize}

This setup enables an empirical comparison between model-generated outputs and real patient language patterns, thereby validating the functional alignment of the simulated impairments with clinical language disorders.

\begin{figure*}[t]
    \centering
    \includegraphics[width=0.95\textwidth]{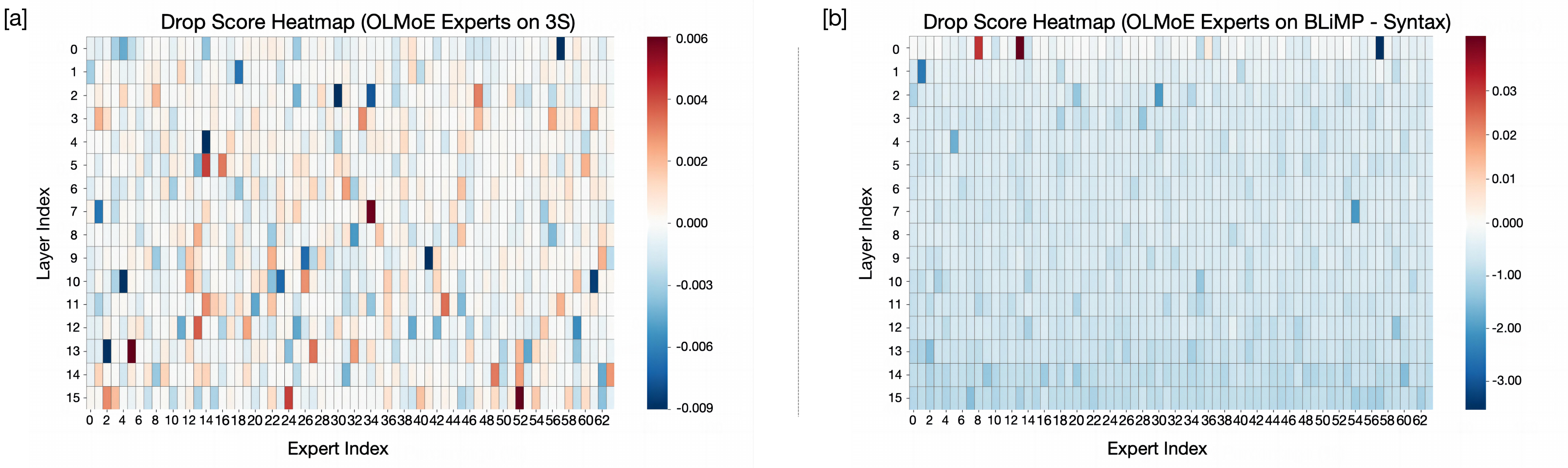}
    \caption{
    The heatmaps visualize the performance change on [a] semantic (3S) and [b] syntactic (BLIMP) benchmarks after lesioning each expert individually. Blue indicates a performance drop, signifying the expert's importance for the task, while red indicates a performance increase. Axes represent model layers (vertical) and expert indices (horizontal).
    }
    \label{fig:fig3}
\end{figure*}

\subsection{Retraining to simulate recovery}

In clinical settings, language therapy—including oral production, comprehension training, and conversational practice—is widely recognized as an effective intervention for aphasia recovery \citep{bhogal2003intensity, stefaniak2022multidimensional}.  Drawing inspiration from these therapeutic strategies, we simulate recovery by re-exposing the damaged model to natural dialogue data, aiming to replicate the iterative process of understanding, responding, and refining that characterizes human rehabilitation.

Specifically, we employed two distinct datasets to retrain our model, aiming to simulate the rehabilitation process for aphasia patients. First, we used the DailyDiaLog dataset to emulate conversational training commonly applied in clinical settings. This dataset was segmented into subsets representing 20\%, 50\%, 75\%, and 100\% of the total data, enabling us to systematically assess the model's recovery progression. Second, specialized datasets were selected to target specific linguistic deficits: the 3S (SICK + STSB + STS12)  dataset was used to facilitate semantic training tailored for Wernicke’s aphasia, while the syntax-specific subset of the BLiMP dataset was employed for syntactic rehabilitation relevant to Broca’s aphasia. Both specialized datasets were divided into training (70\%) and post-training evaluation (30\%) sets.

Prior to fine-tuning, we froze the parameters associated with the damaged experts within the model, restricting updates exclusively to the undamaged experts. This approach mirrors the clinical scenario where damaged neural areas in aphasia patients remain impaired, and rehabilitation is achieved through compensatory mechanisms in other neural regions. Following retraining, we evaluated the model performance using the original benchmark tasks (3S and BLiMP-Syntax datasets) and analyzed the regenerated text responses to standardized prompts derived from AphasiaBank. The goal is to observe:
\begin{itemize}
    \item Improvements in task performance metrics
    \item Qualitative restoration in generated language, such as increased fluency, reduced semantic errors (for Wernicke), and more structured grammar (for Broca).
\end{itemize}
This phase enables us to investigate whether functional impairments in modular language models are reversible through targeted retraining, thereby drawing parallels with therapeutic interventions in human aphasia recovery.

\section{Experimental Setup}
\subsection{Dataset}
In this paper, we use four datasets:
\begin{itemize}
\item \textbf{BLiMP - Syntax:} The Benchmark of Linguistic Minimal Pairs \citep{warstadt2020blimp} tests grammatical knowledge via minimal sentence pairs. We selected 26 syntax-focused subsets (26,000 samples) to form BLiMP-Syntax for evaluating syntactic ability.
\item \textbf{3S (SICK + STSB + STS12) Dataset:} We combined three semantic similarity benchmarks—SICK \citep{marelli2014semeval}, STSB \citep{cer2017semeval}, and STS12 \citep{agirre2016semeval}—into a unified 3S dataset. SICK (10K samples) focuses on compositional semantics, STSB (5.7K) covers diverse real-world sources, and STS12 (2.2K) derives from early SemEval tasks. After merging and filtering, the final 3S dataset contains 17,824 sentence pairs for evaluating semantic understanding.
\item \textbf{Aphasia Bank:} Aphasia Bank \citep{macwhinney2011aphasiabank} contains conversations, narratives, and Q\&A sessions involving individuals with aphasia. To simulate Wernicke’s and Broca’s aphasia, we extracted 63 Wernicke’s and 313 Broca’s aphasia samples, excluding image-based tasks, for controlled evaluation of our aphasia-simulated models.
\item \textbf{DailyDiaLog:} We use the DailyDiaLog dataset \citep{li2017dailydialog} as a training dataset for model recovery. The dialogues in DailyDialog simulate daily written English conversations, covering interpersonal communication, daily events, emotional exchanges, etc.
\end{itemize}

\begin{figure*}[t]
    \centering
    \includegraphics[width=0.95\textwidth]{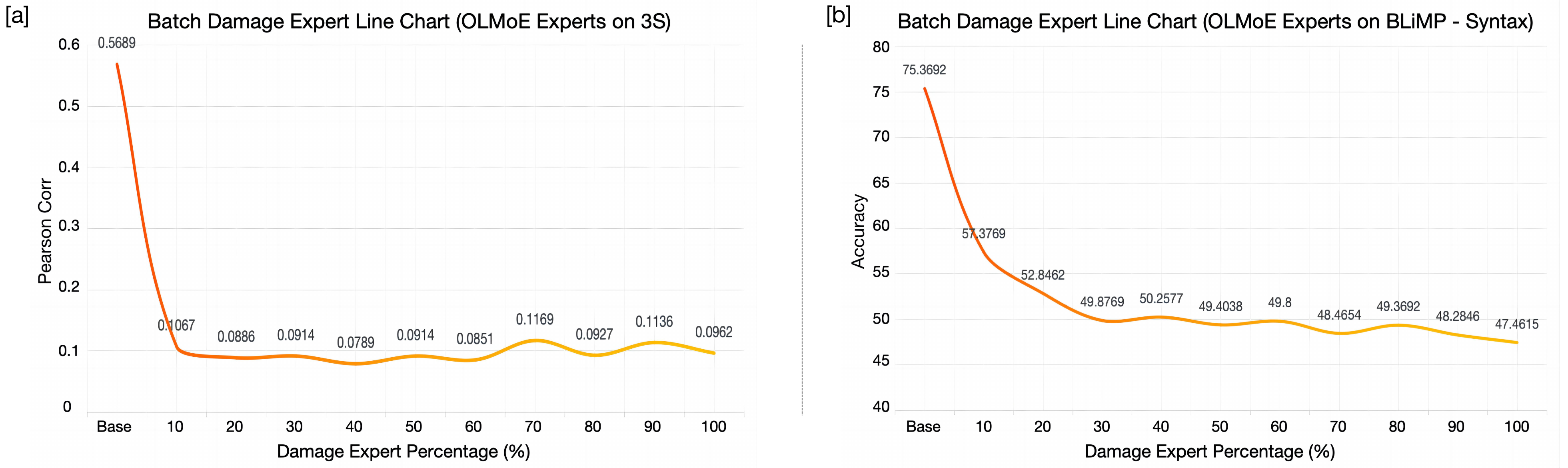}
    \caption{\textbf{Model performance under increasing levels of damage}. The line charts show model scores for [a] semantic and [b] syntactic tasks as the percentage of lesioned experts increases from 10\% to 100\%, compared to the undamaged baseline ("Base").
    }
    \label{fig:fig4}
\end{figure*}

\begin{figure*}[t]
    \centering
    \includegraphics[width=0.95\textwidth]{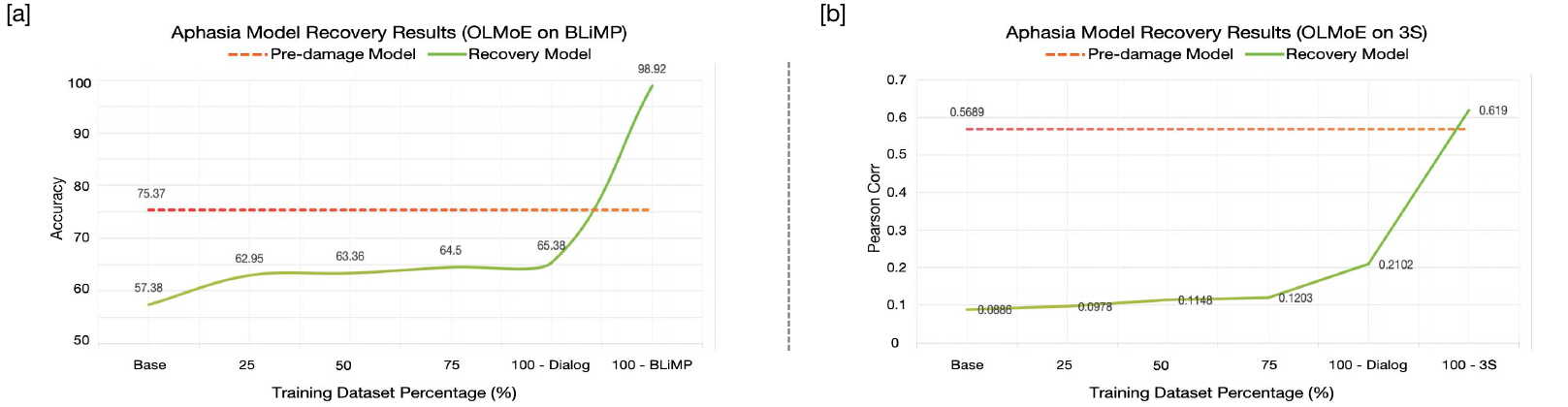}
    \caption{\textbf{Model performance under gradual recovery}. The line graphs show the model scores on [a] syntactic tasks and [b] semantic tasks as the percentage of DailyDialog on the training dataset increases from 25\% to 100\%, compared to the unimpaired baseline ("Red line (Pre-impairment model)"). '100-3S' and '100-BLiMP' are the scores on the semantic and syntactic training datasets.
    }
    \label{fig:fig5}
\end{figure*}

\subsection{Evaluation Metrics}
During model damage evaluation, we adopt two metrics corresponding to two dataset types. For the 3S dataset (SICK, STSB, STS12), we compute the Pearson correlation coefficient between model-predicted and human-annotated semantic similarity scores, assessing the model’s ability to capture sentence-level meaning alignment. A higher coefficient indicates better semantic sensitivity.

For the BLiMP-Syntax dataset, we compute accuracy based on average log probability. Each sample contains a grammatically correct and incorrect sentence. If the model assigns higher average log probability to the correct sentence, it is counted as accurate. This metric reflects the model’s syntactic judgment ability.

\subsection{Implementation Details}
For single- and batch-expert ablation, we performed inference only, using DeepSpeed for distributed inference and memory optimization on four Tesla V100 32GB GPUs with a batch size of 256. During recovery, training was conducted on four H20 NVLink 96GB GPUs using DeepSpeed, with a learning rate of 5e-5 and the AdamW optimizer. The semantic-damage model was trained for 3 epochs, and the syntactic-damage model for 1 epoch.

\section{Results and Discussion}

This section presents the full experimental results. We begin by analyzing the impact of single-expert impairments. We then simulate aphasia via cumulative batch damage and examine the resulting text generation. Finally, we assess model recovery using targeted retraining and evaluate restoration of linguistic function.

\begin{table*}[t]
\renewcommand{\arraystretch}{1.2}
\begin{adjustbox}{width=1\textwidth,center}
\begin{tabular}{cccccccccccccc}
\hline
\multicolumn{1}{l}{} & \multicolumn{6}{c}{\textbf{OLMoE, 3S (SICK, STS-B, STS12)}} & \multicolumn{1}{l}{} & \multicolumn{6}{c}{\textbf{OLMoE, BLiMP - Syntax}} \\
\cline{2-7} \cline{9-14}

  & \textbf{Dataset} & \textbf{\makecell{Pearson\\Corr}} & \textbf{Base} & \textbf{\makecell{Layer\\Index}} & \textbf{\makecell{Expert\\Index}} & \textbf{\makecell{Drop\\Score}} &   & \textbf{Dataset} & \textbf{Accuracy} & \textbf{Base} & \textbf{\makecell{Layer\\Index}} & \textbf{\makecell{Expert\\Index}} & \textbf{\makecell{Drop\\Score}} \\
\cline{1-7} \cline{9-14}
1 & \cellcolor{lightred}3S & \cellcolor{lightred}0.447 & \cellcolor{lightred}0.5689 & \cellcolor{lightred}0 & \cellcolor{lightred}57 & \cellcolor{lightred}-0.1219 & & \cellcolor{lightred}BLiMP & \cellcolor{lightred}71.8269 & \cellcolor{lightred}75.3692 & \cellcolor{lightred}0 & \cellcolor{lightred}57 & \cellcolor{lightred}-3.5423  \\
\cline{1-7} \cline{9-14}
2 & \cellcolor{lightblue}3S & \cellcolor{lightblue}0.5364 & \cellcolor{lightblue}0.5689 & \cellcolor{lightblue}2 & \cellcolor{lightblue}30 & \cellcolor{lightblue}-0.0325 & & BLiMP & 72.8577 & 75.3692 & 1 & 1 & -2.5115  \\
\cline{1-7} \cline{9-14}
3 & \cellcolor{lightgreen}3S & \cellcolor{lightgreen}0.5517 & \cellcolor{lightgreen}0.5689 & \cellcolor{lightgreen}13 & \cellcolor{lightgreen}2 & \cellcolor{lightgreen}-0.0172 & & BLiMP & 73.3346 & 75.3692 & 7 & 54 & -2.0346  \\
\cline{1-7} \cline{9-14}
4 & 3S & 0.5558 & 0.5689 & 4 & 14 & -0.0131 & & \cellcolor{lightblue}BLiMP & \cellcolor{lightblue}73.3577 & \cellcolor{lightblue}75.3692 & \cellcolor{lightblue}2 & \cellcolor{lightblue}30 & \cellcolor{lightblue}-2.0115  \\
\cline{1-7} \cline{9-14}
5 & 3S & 0.5588 & 0.5689 & 9 & 41 & -0.0101 & & BLiMP & 73.6692 & 75.3692 & 4 & 5 & -1.7  \\
\cline{1-7} \cline{9-14}
6 & 3S & 0.56 & 0.5689 & 10 & 4 & -0.0089 & & BLiMP & 73.8654 & 75.3692 & 14 & 60 & -1.5038  \\
\cline{1-7} \cline{9-14}
7 & 3S & 0.5604 & 0.5689 & 10 & 61 & -0.0085 & & \cellcolor{lightgreen}BLiMP & \cellcolor{lightgreen}73.8885 & \cellcolor{lightgreen}75.3692 & \cellcolor{lightgreen}13 & \cellcolor{lightgreen}2 & \cellcolor{lightgreen}-1.4807  \\
\cline{1-7} \cline{9-14}
8 & 3S & 0.5613 & 0.5689 & 2 & 34 & -0.0076 & & BLiMP & 73.9192 & 75.3692 & 15 & 7 & -1.45  \\
\cline{1-7} \cline{9-14}
9 & 3S & 0.5618 & 0.5689 & 10 & 23 & -0.0071 & & \cellcolor{lightyellow}BLiMP & \cellcolor{lightyellow}73.9769 & \cellcolor{lightyellow}75.3692 & \cellcolor{lightyellow}12 & \cellcolor{lightyellow}59 & \cellcolor{lightyellow}-1.3923  \\
\cline{1-7} \cline{9-14}
10 & 3S & 0.562 & 0.5689 & 9 & 26 & -0.0069 & & BLiMP & 74.0192 & 75.3692 & 2 & 20 & -1.35  \\
\cline{1-7} \cline{9-14}
11 & 3S & 0.5623 & 0.5689 & 1 & 18 & -0.0066 & & BLiMP & 74.0192 & 75.3692 & 3 & 28 & -1.35  \\
\cline{1-7} \cline{9-14}
12 & 3S & 0.5625 & 0.5689 & 7 & 1 & -0.0064 & & BLiMP & 74.0346 & 75.3692 & 14 & 9 & -1.3346  \\
\cline{1-7} \cline{9-14}
13 & \cellcolor{lightyellow}3S & \cellcolor{lightyellow}0.5633 & \cellcolor{lightyellow}0.5689 & \cellcolor{lightyellow}12 & \cellcolor{lightyellow}59 & \cellcolor{lightyellow}-0.0056 & & BLiMP & 74.0385 & 75.3692 & 10 & 35 & -1.3307  \\
\cline{1-7} \cline{9-14}
14 & 3S & 0.5638 & 0.5689 & 8 & 32 & -0.0051 & & BLiMP & 74.0962 & 75.3692 & 14 & 16 & -1.273  \\
\cline{1-7} \cline{9-14}
15 & 3S & 0.5643 & 0.5689 & 11 & 20 & -0.0046 & & BLiMP & 74.1385 & 75.3692 & 14 & 20 & -1.2307 \\
\hline
\end{tabular}
\end{adjustbox}
\caption{\textbf{Impairment of Single Expert Results Table.} This table summarizes the single-expert impairment results for the semantic impairment model (left) and the syntactic impairment model (right). “Base” represents the performance of the unimpaired model, and “Drop Score” denotes the performance decrease after impairing a single expert. The top 15 most impactful experts are listed. Experts highlighted in the same color significantly affect both semantic and syntactic impairment tasks.}
\label{tab:1}
\end{table*}

\subsection{Impairment of Single Expert}
To analyze expert function and to lay the groundwork for bulk lesion experiments, we simulated aphasia by selectively impairing the abilities of individual experts. Figure \ref{fig:fig3} visualizes these results via Drop Score heatmaps. In total, 551 experts in the semantic tasks (SICK, STS-B, STS12) and 1,020 experts in the syntactic task (BLiMP-Syntax) caused a performance drop when individually ablated.

As shown in the Drop Score heatmaps (Figure \ref{fig:fig3}a for semantic tasks and Figure \ref{fig:fig3}b for syntactic tasks), we observe that impairing certain individual experts leads to consistent variations in model performance. While a small number of experts cause noticeable drops, most exhibit relatively minor yet consistent effects. This pattern holds across both semantic and syntactic evaluations. Importantly, the presence of these stable, low-magnitude effects suggests a distributed contribution across the expert modules, and supports the hypothesis that selective impairment—even of less impactful experts—can subtly alter model behavior. These findings motivate a more systematic investigation of expert interaction and redundancy, which we pursue in the following section through progressive batch impairment experiments.

Comparing both tasks, we found overlap among experts with the most severe Drop Scores. Among the top 15 highest-impact experts for both tasks, 4 were shared ((0, 57), (2, 30), (13, 2) and (12, 59)) (Summarized in Table \ref{tab:1}). These general-purpose experts are critical to both semantic and syntactic functions, acting as multi-task hubs. To ensure our simulations reflect specific language impairments, we exclude these core experts from batch ablation, enabling targeted modeling of Broca’s or Wernicke’s aphasia.

\begin{table*}[t]
    \centering
    \includegraphics[width=0.9\textwidth]{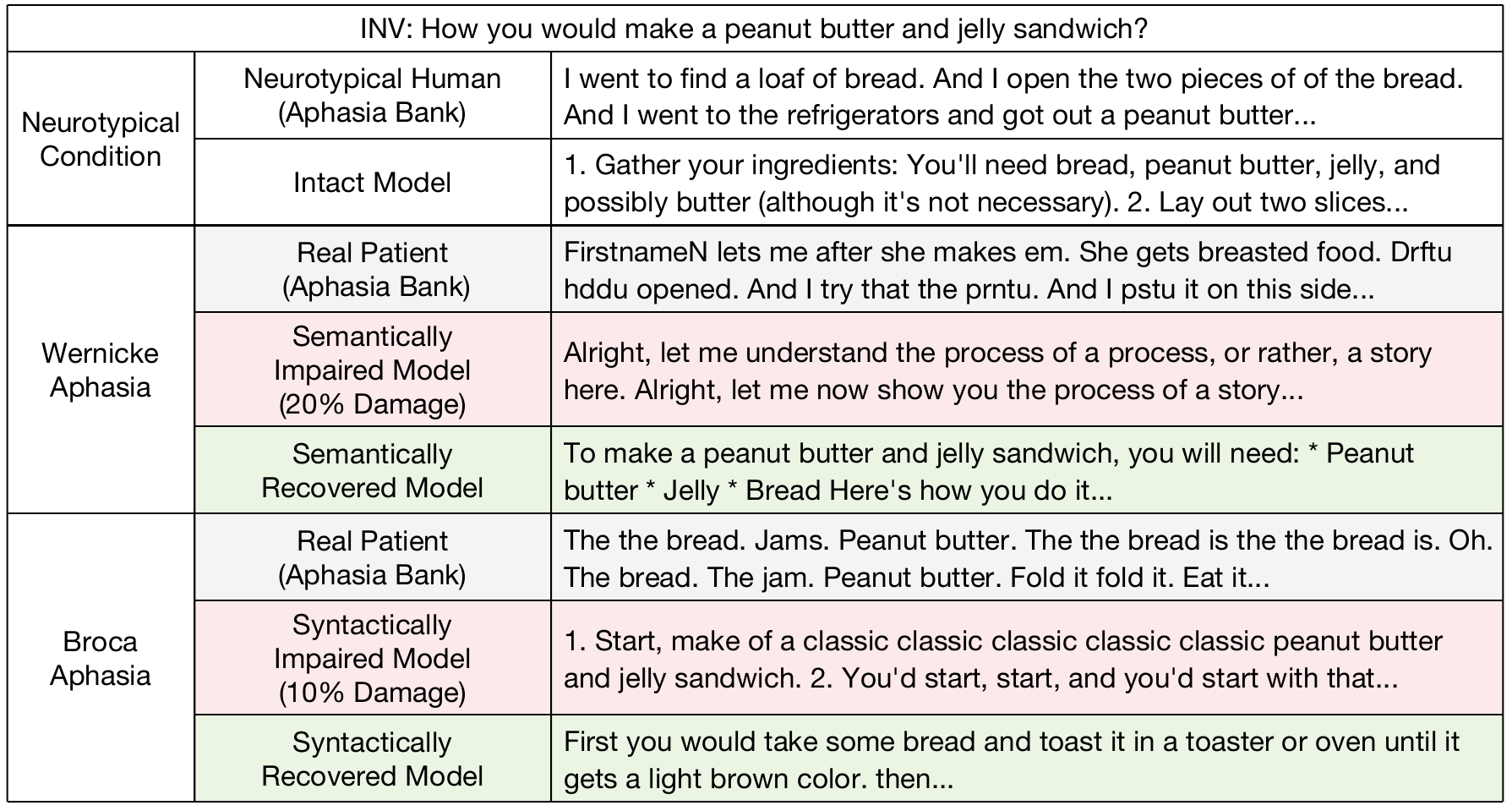}
    \caption{\textbf{Comparison of textual responses to a descriptive prompt}: Outputs are shown for both human speakers (neurotypical, Wernicke's, and Broca's aphasia patients from AphasiaBank) and our model under different conditions: intact, impaired (semantic or syntactic), and post-recovery. The prompt (INV) is from the original AphasiaBank interview.}
    \label{tab:2}
\end{table*}

\subsection{Batch Damage and Aphasia Text Generation}
Building on the single-expert impairment results, we conducted batch ablation experiments by cumulatively impairing experts ranked by Drop Score. Experts with the most negative impact were grouped into 10\% to 100\% impairment tiers. Figure \ref{fig:fig4} presents performance trends across semantic and syntactic benchmarks.

For semantic tasks (Figure \ref{fig:fig4}a), performance dropped sharply from a baseline of 0.5689 to 0.1067 after impairing the top 10\% of experts, revealing that \textbf{a small set of experts are critical for semantic understanding.} As more experts were impaired, performance stabilized around 0.1, indicating redundancy among remaining experts but limited compensatory capacity.

For syntactic tasks (Figure \ref{fig:fig4}b), accuracy declined from 75.37\% to 57.38\% at 10\% impairment and gradually dropped to ~49\% beyond 30\% impairment. The slower decline suggests mild redundancy but persistent syntactic degradation, reinforcing that \textbf{syntax depends on a smaller subset of specialized experts.} Combined with the results of single expert impairments on the previous two tasks, while experts with low Drop Scores can degrade model performance, the impact is minimal. This also explains why, in the batch impairment experiments, the scores for both tasks stabilize as the number of impaired experts increases, starting at 30\%.

To validate our simulations, we compared model outputs to real patient responses from AphasiaBank for the prompt "How would you make a peanut butter and jelly sandwich?". As shown in Table \ref{tab:2}, Wernicke’s patient is characterized by fluent and grammatically well-formed speech that is often semantically confused or irrelevant. For example, the patient utterance “she gets breasted food” follows correct syntax but is nonsensical in meaning. Our semantically impaired model exhibits similar behavior, generating “understand the process of a process”, which maintains fluency but lacks coherent content. Broca’s patient involves non-fluent, fragmented speech with frequent repetition and disrupted syntax. The patient output “The the bread. Jams. Peanut butter...The the bread. The jam. Peanut butter…” reflects halted sentence and poor structure. Likewise, the syntactically impaired model produces “make of a classic classic classic classic peanut butter and jelly sandwich”, showing repetitive phrasing and disrupted syntax.

These findings demonstrate that \textbf{targeted expert damage in OLMoE reliably mimics clinical aphasia patterns. Beyond certain thresholds} (20\% semantic, 10\% syntactic), \textbf{models begin producing incoherent text}, establishing optimal impairment levels for subsequent recovery experiments on functional compensation and rehabilitation.

\subsection{Aphasia Model Recovery Results}
Following identification of optimal lesion thresholds (20\% semantic, 10\% syntactic), we retrained the impaired models using progressively larger subsets of the DailyDialog dataset. Recovery was evaluated on subtype-specific benchmarks. Figure \ref{fig:fig5} shows the performance trends of models trained with different proportions of the training set evaluated on the benchmark.

For the Broca model (Figure \ref{fig:fig5}a), the initial accuracy was 57.38\%. As the training data gradually increased from 25\% to 100\% of the DailyDialog corpus, the accuracy steadily improved to 62.95\%, 63.36\%, 64.50\%, and finally to 65.38\%, showing consistent gains in syntactic performance. Although still below the original 75.37\% baseline, this demonstrates partial recovery through conversational training. In comparison, the Wernicke model achieved an initial score of only 0.0886 (Figure \ref{fig:fig5}b). After retraining on the same incremental subset of DailyDialog data, the model achieved scores of 0.0978, 0.1148, 0.1203, and finally 0.2102. Reflecting notable though limited improvement in semantic capabilities compared to its baseline of 0.5689. To further validate recovery, we conducted targeted retraining on subtype-specific datasets (3S for Wernicke, BLiMP-Syntax for Broca), using 70/30 splits and 5-fold cross-validation. The Wernicke model reached 0.619, exceeding its baseline, while the Broca model achieved 98.92\%, substantially outperforming its original score. \textbf{These results highlight the importance of tailored training in restoring function.}

In summary, \textbf{targeted retraining of the undamaged components in the MoE model significantly restores linguistic abilities.} This supports modular LLMs as a promising framework for modeling and rehabilitating language impairments through focused multi-task training.

\section{Conclusion and Future Work}

We began this work with a central question: can the modular architecture of Mixture-of-Experts (MoE) models serve as a computational proxy for the brain's functional specialization, allowing us to simulate and rehabilitate language disorders? Our findings provide a strong affirmative answer. By selectively lesioning experts, we successfully replicated the distinct linguistic profiles of Broca’s and Wernicke’s aphasia, inducing targeted performance drops on syntactic and semantic benchmarks (up to 23.9\% and 84.4\%, respectively) and producing outputs that qualitatively align with patient speech from AphasiaBank. Crucially, we showed that this computational model is not merely static; in a process analogous to clinical therapy, we froze the damaged experts and retrained the intact network on conversational data, leading to a robust recovery of linguistic function of over 40\%. This dual achievement—in successfully modeling both impairment and rehabilitation—establishes modular LLMs as a powerful new framework for computational neuropsychology, offering a viable platform for investigating the mechanisms of language disorders and for prototyping novel, data-driven therapeutic interventions.

Future work will extend this framework to more complex conversational tasks and other aphasia subtypes, such as conduction aphasia. Key directions include investigating the underlying mechanisms of network plasticity during recovery and expanding our methodology to multilingual and multimodal contexts. Ultimately, this research path aims to leverage computational models to design and test personalized, scalable therapeutic interventions.

\section{Acknowledgements}
This research was supported by grants from the National Natural Science Foundation of China to S.W. (62036001) and S.W. (the STI2030-Major Project, grant number: 2021ZD0204105).

\bibliography{main}
\end{document}